\documentclass[11pt,oneside]{article} 
\usepackage[margin=1in]{geometry}

\def\bSig\mathbf{\Sigma}

\usepackage[flushleft]{threeparttable}

\usepackage{amsfonts}
\usepackage{amssymb}
\usepackage{mathtools}
\usepackage{amsmath}
\usepackage[table]{xcolor}
\usepackage{url}

\newtheorem{theorem}{Theorem} 

\usepackage[authoryear]{natbib}
\setcitestyle{authoryear}

\usepackage{authblk}

\newcommand{\independent}{\perp\!\!\!\!\perp} 

\begin{document}


\title{Nonparametric Estimation of Event-Free Survival for Data with Left-Truncated Death and Intermittently Assessed Nonfatal Events}

\author[1]{Han Lu\thanks{\textit{email:} lu000054@umn.edu}}
\author[1]{Xianghua Luo}
\author[2]{Yifei Sun}
\author[3,4]{Wendy Wang}
\author[5]{Thomas Mosley}
\author[6]{Priya Palta}
\author[7]{Elsayed Z. Soliman}
\author[8]{Lin Yee Chen}
\author[1]{Anne Eaton\thanks{\textit{email:} eato0055@umn.edu}}

\affil[1]{Division of Biostatistics and Health Data Science, School of Public Health, University of Minnesota}
\affil[2]{Department of Biostatistics, Mailman School of Public Health, Columbia University}
\affil[3]{Department of Epidemiology, Peter O’Donnell Jr. School of Public Health, University of Texas}
\affil[4]{Department of Internal Medicine, University of Texas Southwestern Medical Center}
\affil[5]{The MIND Center, University of Mississippi Medical Center}
\affil[6]{Department of Neurology, School of Medicine, University of North Carolina}
\affil[7]{Epidemiological Cardiology Research Center, Department of Cardiovascular Medicine, Wake Forest University School of Medicine}
\affil[8]{Department of Medicine, Medical School, University of Minnesota}

\date{}

\maketitle

\begin{abstract} 

In certain clinical settings, patients diagnosed with a disease of interest are at risk for death as well as a serious nonfatal event, and interest lies in event-free survival (EFS), a composite endpoint defined as the time from disease onset until the earlier of the nonfatal event and death. Component-wise censoring of EFS arises when each component is subject to a different censoring mechanism. For example, the nonfatal event may be interval censored between assessments, and death is right-censored. Further, in studies where individuals enroll after disease onset (including prevalent cohort studies), EFS is left truncated. Methods to estimate EFS probability with left-truncated and right-censored data are available in the literature, but they cannot handle component-wise censoring. We propose a kernel smoothing method to non-parametrically estimate EFS in this setting. Our method can also estimate and test for differences in the restricted mean event-free survival time, and can leverage two types of supplemental data that may be available: data from participants followed for death only (not followed for the nonfatal event), and incident cohort data, which arises when there is no delay between disease onset and study enrollment. We assess the proposed method using simulations and demonstrate the method using data from the Atherosclerosis Risk in Communities (ARIC) Study to estimate dementia-free survival probability following a myocardial infarction.

\end{abstract}

\maketitle

\section{Introduction}
\label{s:intro}

In biomedical research, interest often lies in multiple time-to-event endpoints. Composite endpoints defined as the time from an index event to either a nonfatal event of interest or death, whichever occurs first, are common; we refer to such endpoints as event-free survival (EFS). EFS is of high clinical interest because it reflects disease progression, not just death, and the use of EFS as a primary endpoint can potentially shorten the follow-up time required in a clinical trial, compared to overall survival. For example, in oncology, progression-free survival is frequently considered as a primary endpoint in drug development \citep{carroll_2007}. 

A conventional approach to estimate EFS is the Kaplan-Meier (KM) estimator, which assumes that both the nonfatal event time and death time are right-censored. However, this assumption is often violated in practice, as in many studies, the nonfatal event is assessed periodically (for example, at clinic visits), resulting in interval censoring. Meanwhile, the exact time of death is right-censored. Since each component of the composite endpoint is subject to a different censoring mechanism, this setting leads to what we call ``component-wise censoring'' \citep{eaton_etal_2022}. By treating nonfatal events as if they occurred on the day they were detected, methods designed for right-censored data, such as the KM estimator and the Cox proportional hazards model, can be used on component-wise censored data, but this introduces bias, with the degree of bias depending on the visit schedule, among other factors (\citeauthor{panageas_etal_2007}, \citeyear{panageas_etal_2007}; \citeauthor{zeng_etal_2015}, \citeyear{zeng_etal_2015}; \citeauthor{eaton_zabor_2022}, \citeyear{eaton_zabor_2022}; Speiser, Ambrosius, and Pajewski, 2023). To address component-wise censoring, several alternative estimation approaches have been developed. \citet{gao_etal_2025} proposed to address component-wise censoring by treating the right-censored component as interval-censored and applying a conventional interval censoring framework. Based on a proportional hazards model, \citet{diao_etal_2018} proposed a semiparametric likelihood-based method for component-wise censored data. \citet{eaton_etal_2022} proposed a nonparametric kernel estimator of EFS under component-wise censoring. 
A review of available estimators for component-wise censored data, including the naive KM estimator, the kernel method, a nonparametric maximum likelihood estimator, and a parametric estimator, is provided by \citet{eaton_2024}. 

A limitation of existing methods for component-wise censored data is that they were developed for use on incident cohort data; hence, they are not appropriate for data with left truncation, which can arise in multiple settings, including prevalent cohort studies, studies that use electronic health record data or registry data, and studies where eligibility requires reaching a specified age. For left-truncated data without component-wise censoring, product-limit estimators can be used to estimate EFS. Specifically, left truncation without censoring can be addressed with the Lynden-Bell estimator \citep{lynden_bell_1971}, and left truncation with right-censoring can be addressed using a product-limit estimator similar to the KM estimator, which adjusts risk sets to account for delayed entry (\citeauthor{turnbull_1976}, \citeyear{turnbull_1976}; Tsai, Jewell, and Wang, 1987). This product-limit estimator is not intended for left-truncated and component-wise censored data, and its use for such data requires treating the component-wise censored endpoint as right-censored (e.g., treating nonfatal events as if they occurred on the day they were detected). This highlights the need for methods that simultaneously address both left truncation and component-wise censoring. 

In this paper, we propose nonparametric kernel-based estimators for EFS probability and its associated event-free restricted mean survival time (EF-RMST) to handle data with left-truncated death and component-wise censoring in a prevalent cohort. Our method can handle a mix of incident and prevalent cohort data, and we investigate the advantages of incorporating incident cohort data, similar to \citet{wolfson_etal_2019}. The proposed method can also handle study designs where some participants are followed for both the nonfatal event and death, and some participants are followed for death only. The performance of the proposed method and comparison methods is evaluated with simulations. We apply the proposed method to the Atherosclerosis Risk in Communities (ARIC) Study to estimate dementia-free survival probability and restricted mean dementia-free survival time following myocardial infarction (MI). ARIC is a long-running community-based cohort study that has followed participants since 1987 \citep{wright_etal_2021}, with MI adjudicated since Visit 1 (1987 to 1989) and dementia assessed intermittently since Visit 5 (2011 to 2013). To estimate dementia-free survival, we include ARIC participants who were alive at Visit 5. Participants who had their first MI between Visit 1 and Visit 5 and remained alive at Visit 5 are prevalent cases, while participants who had their first MI after Visit 5 are incident cases. 

\section{Methods}
\label{s:methods}

\subsection{Prevalent cohort and supplemental cohorts} \label{s:Def PC IC}

\begin{figure}
\centering
\caption{\label{fig:prev_inci} Possible observations in prevalent and incident cohorts under calendar time scale (left panel) and personal time scale (right panel). On a calendar scale, $s_1$ is the theoretical earliest day to observe an index event, $v$ is the prevalence day, and $\tau_{eos}$ is the end of the study. On a personal scale, time 0 starts with the occurrence of the index event, and $\tau=\tau_{eos}-s_1$ is the theoretical longest follow-up time for a participant. 
}
\includegraphics[width=1\linewidth]{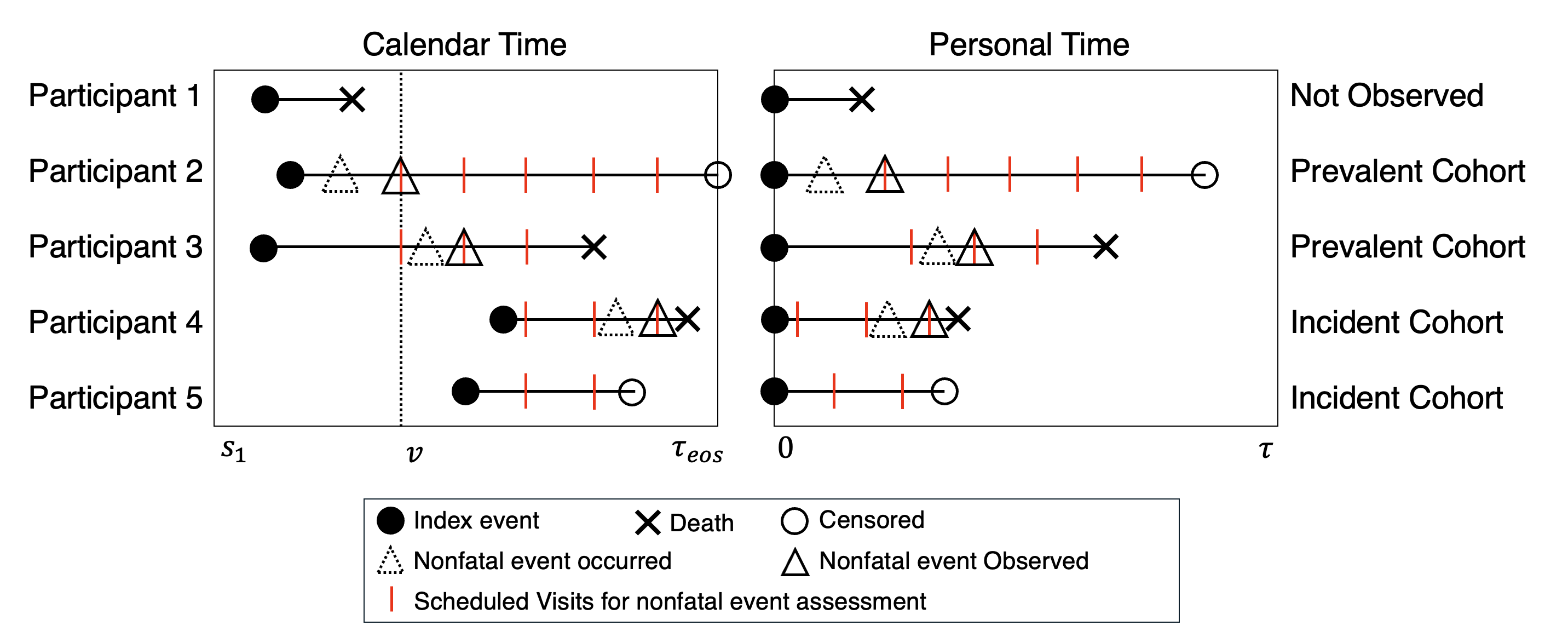}
\end{figure}

We define the prevalent and incident cohorts in a longitudinal study framework following \citep{wolfson_etal_2019}, where participants are recruited and screened for the index event on a prevalence day $v$. Those who have experienced the index event (e.g., MI in ARIC) prior to $v$ and are alive at $v$ are enrolled in the prevalent cohort, followed for the nonfatal event (e.g., dementia in ARIC) and death, regardless of whether they have had the nonfatal event by $v$. Note that this definition is used for simplicity and that in practice, participants would typically be recruited and screened over a period of time. The theoretical earliest date to observe an index event is denoted by $s_1$, and $\tau_{eos}$ is the calendar date for the end of the study. Then, the theoretical longest follow-up time for this study is $\tau = \tau_{eos}-s_1$. A visualization of the study timeline and possible participants is shown in Figure \ref{fig:prev_inci}. Individuals who had already experienced the index event by $v$ and who were alive at $v$ are included in the prevalent cohort, such as participants 2 and 3 in Figure \ref{fig:prev_inci}. Their left truncation times are defined as the time between the index event and day $v$. Note that participant 1 is not observed due to left truncation. 

Those who have not experienced the index event on the prevalence day $v$ are enrolled in the incident cohort, followed for the index event, nonfatal event, and death, such as participants 4 and 5 in Figure \ref{fig:prev_inci}. While we aim to propose a method for the prevalent cohort, since the incident cohort and the prevalent cohort arise from the same target population, the incident cohort can be used to supplement the prevalent cohort data. Both \citet{wolfson_etal_2019} and \citet{lee_etal_2019} had demonstrated the statistical benefits of jointly analyzing the combined cohort. The data are collected in calendar time and can be converted to personal time anchored at the index event, as shown in the right panel of Figure \ref{fig:prev_inci}. 

We define the death-only cohort as distinct from the incident and prevalent cohorts, including those who, by design, have no follow-up visits for the nonfatal event and are only followed for death. A death-only cohort could arise from an external source (e.g., National Death Index) or study designs where only a subset of the participants are followed for the nonfatal event. The death-only cohort can be used to supplement the prevalent cohort data as well, when they are from the same target population. Participants who had no events observed at visits (e.g., participant 5 in Figure \ref{fig:prev_inci}), who missed visits, or who passed away before any visits could occur do not belong to the death-only cohort. The death-only cohort can be further divided into the death-only prevalent cohort and the death-only incident cohort based on the timing of the index event relative to the prevalence day.

\subsection{Event times and visit schedule in a prevalent cohort} \label{s:Def event times}

Subsequently, all times are defined on the personal time scale (i.e., time since the index event). Notation and estimators will be developed for the prevalent cohort setting, and supplementation with incident cohort data will be discussed in Section \ref{s:proposed supp}.

In the population from which the prevalent cohort arises, let $D^0$ be the population time from the index event to death, and let $L^0$ be the population time from the index event to study enrollment on the prevalence day. Let $Y^0(t)$ be the nonfatal-event-free indicator at time $t$ after the index event for $t \geq 0$; $Y^0(t)=1$  if a participant has not yet had a nonfatal event by time $t$ and $Y^0(t)=0$ if they have. In a prevalent cohort, participants can only be observed if $D^0>L^0$. Under the biased sampling arising from left truncation, samples $\{ D, L, Y(\cdot) \}$ have the same distribution as the population $\{ D^0, L^0, Y^0(\cdot) \}$ given $D^0 > L^0$. The sampled death time, $D$, is subject to the corresponding sampled left truncation time $L$, and right-censoring time $C=L+C_{res}$, where $C_{res}$ is the residual censoring time (i.e., the censoring time from study enrollment). Define $X = \min(D, C)$ and $\Delta = I(D \leq C)$. Then, $Y(t)$ is the sampled nonfatal-event-free indicator at time $t$.

Define the potential visit counting process on and after the prevalence day $N^*(s)$ for $s\geq 0$. In practice, with participants enrolled over a period of time, this means that each participant's potential visit counting process is defined with respect to their study enrollment time (i.e., $s=0$ is study enrollment time). The occurrence of observed visits can be expressed as $dN(t) = dN^*(t-L)I(t\leq X)$ for $t \geq L$, and $dN(t)=0$ for $t<L$. Therefore, observed visits are a subset of potential visits. A patient's nonfatal event status can only be assessed at a visit; that is, $Y(t)$ is only observed at the times for which $N(t)$ increases. Note that study entry depends on being alive but not on nonfatal event status $Y(t)$. Participants who experienced the non-fatal event but are still alive at time $L$ are included in the study. The rate function of the observed visit process is $\lambda(t) dt = E\{ dN(t) \}$. We assume that $\lambda(t)$ is positive on $t \in [0, \tau]$. 

The observed data are $\{L_i, X_i, \Delta_i, Y_i(t)dN_i(t)\}$ for $0 \leq t \leq \tau$ and $ i=1, \cdots, n$. We assume that $L^0 \independent \{D^0, Y^0(\cdot) \}$, generalizing the commonly used independent left truncation assumption to include the nonfatal event-free marker as well. That is, at the population level, the delayed entries are unrelated to how long a participant will be alive or free of the nonfatal event (no cohort effect). We also assume that $\{C_{res}, N^*(\cdot)\} \independent \{D, Y(\cdot) \}|L$, meaning that conditional on the left truncation time, the censoring and visit processes are independent of the timing of the nonfatal event and death. Note that the conditional independent visit process assumption is reasonable for planned visit schedules, such as in a clinical trial or prospective cohort study. However, this assumption can be violated if participants have visits due to symptoms related to the nonfatal event or death. 

\subsection{Proposed estimators for prevalent cohort data}\label{s:proposed prev}

In \citet{eaton_etal_2022}, the estimator for the EFS probability in an incident cohort is based on decomposing EFS into the product of two probabilities. We follow a similar decomposition and propose to modify the estimation of each component in order to extend the method to left-truncated data. For $t \in [0,\tau]$, EFS at time $t$ can be expressed as $p(t) = E[Y^0(t)I(D^0\geq t)] = E[Y^0(t)|D^0 \geq t]P(D^0\geq t) = r(t)S_{D^0}(t)$, where $S_{D^0}(t) \equiv P(D^0\geq t)$ and $r(t) \equiv E[Y^0(t)|D^0\geq t]$, and we propose to estimate $S_{D^0}(t)$ and $r(t)$ using a product-limit estimator and a kernel estimator, as explained below.

We propose to estimate $S_{D^0}(t)$, the probability of being alive, using the left-truncation,  right-censoring product limit (LT-RC-PL) estimator, following definitions in \citet{tsai_etal_1987}:
\begin{equation} \label{eq:hatsd}
\begin{aligned}
    \hat{S}_D(t) = \prod_{j:t_{(j)}<t} \left(1-\frac{d_j}{a_j}\right)
\end{aligned}
\end{equation}
for $t>t_{(1)}$ and $\hat S_D(t) = 1$ for $t\leq t_{(1)}$. In the estimator above, $t_{(1)}, \cdots, t_{(K)}$ are $K$ distinct ordered death times, $d_j$ is the number of deaths at time $t_{(j)}$, and $a_j = \sum_{i=1}^n I(L_i \leq t_{(j)} \leq X_i)$ is the number of patients at risk at time $t_{(j)}$, where the risk set is modified for left-truncated data. 

We propose to estimate $r(t)$, the probability of being event-free conditional on being alive, by a kernel estimator $\hat{r}_h(t)$ following \citet{eaton_etal_2022} with modifications, where
\begin{equation} \label{eq:hatrh}
\begin{aligned}
    \hat{r}_h(t) &=
        \frac{\sum_{i=1}^n \int_{0}^{\tau} K_{h,t}\left(\frac{t-s}{h}\right) Y_i(s) dN_i(s)}
        {\sum_{i=1}^n \int_{0}^{\tau} K_{h,t}\left(\frac{t-s}{h}\right) dN_i(s)}, \\
\end{aligned}
\end{equation}
for $t \in [0,\tau]$. Here, $K_{h,t}(\cdot)$ is a second-order modified kernel function that has the standard kernel function in the interval $[h, \tau-h]$ and corrects the boundary effects in the intervals $[0,h)$ and $(\tau-h, \tau]$ \citep{muller_1991}. In cases where both the numerator and denominator are zero, we define $\hat r_h(t) = 0$. The bandwidth parameter $h$ controls the smoothness of the kernel estimator $\hat{r}_h(t)$, which drives the bias-variance trade-off.
In summary, the event-free survival probability is estimated as:

\begin{equation} \label{eq:3}
\begin{aligned}
    \hat{p}(t) &= \hat{S}_D(t)\hat{r}_h(t) .\\
\end{aligned}
\end{equation}

Moreover, we propose an estimator for $\mu(t) = \int_0^tp(u)du$, the EF-RMST up to restriction time $t$, $0 < t \leq \tau$ . Unlike the EFS probability curve, which provides a dynamic view of survival over time, the EF-RMST is an interpretable measure that provides a cumulative survival summary up to time $t$. We propose estimating the EF-RMST up to restriction time $t$ as the integral of the above estimated survival probability (equation \ref{eq:3}):

\begin{equation} \label{eq:4}
    \hat\mu(t) = \int_0^{t} \hat{p}(u) du
    = \int_0^{t} \hat{S}_D(u) \hat{r}_h(u) du. \\
\end{equation}

We refer to the proposed estimators in equations \ref{eq:3} and \ref{eq:4} as the LT-Kernel estimators. 

\subsection{Proposed estimators with supplemental data}\label{s:proposed supp}

The proposed estimator shares a drawback with the Lynden-Bell estimator \citep{lynden_bell_1971} and the LT-RC-PL estimator (Wang, Jewell, and Tsai, 1986; \citeauthor{tsai_etal_1987}, \citeyear{tsai_etal_1987}) that estimates can be largely impacted by an event with a very small risk set at early times. This highly variable performance can potentially be addressed by using other adjusted estimators, such as the Lai-Ying estimator \citep{lai_ying_1991} and others (\citeauthor{tsai_1988}, \citeyear{tsai_1988}; \citeauthor{woodroofe_1985}, \citeyear{woodroofe_1985}; \citeauthor{pan_chappell_1999}, \citeyear{pan_chappell_1999}). Another approach, demonstrated by \citet{wolfson_etal_2019} and \citet{lee_etal_2019}, showed the benefits of combining prevalent and incident cohort data. We follow the second approach and propose to incorporate the incident cohort data. We assume that the incident and the prevalent cohort are sampled from the same population with $\{ D^0, L^0, Y^0(\cdot) \}$. The incident cohort can be jointly analyzed with the prevalent cohort by assigning their left truncation time to 0. In addition, the right-censoring time $C$ for the incident cohort is defined as the time from the index event to censoring. 

We additionally propose to supplement the prevalent cohort data with death-only incident or prevalent data, assuming that the supplemental death-only data are also sampled from the same population as the prevalent cohort data. When death-only data is available, while it will not provide any information in estimating ${r} (t)$, and thus the $\hat{r}_h(t)$ remains unchanged, it can be pooled with the existing data to estimate $S_{D^0}(t)$. $\hat{S}_D(t)$ can be estimated by using equation \ref{eq:hatsd} with both prevalent and death-only cohorts data. In particular, $a_j$ becomes the total number of prevalent and death-only participants at risk, resulting in a larger risk set at early times compared to a prevalent-only cohort, and thus improving the performance.

\section{Large Sample Properties}
\label{s:lsp}

The estimator $\hat{\mu}(t)$ is an extension of the estimator developed by \citet{eaton_etal_2022}. The asymptotic properties of $\hat{\mu}(t)$ are derived by making appropriate modifications to the proof in Sun, Huang, and Wang (2017). Define the counting process of observable death $N_i^D(t) = I(X_i \leq t, \Delta_i=1)$ and let $\Lambda_{D^0}(t)$ denote the cumulative hazard of death. Then, the martingale of death is defined as $M_i^D(t) = N_i^D(t) - \int_0^t I(L_i \leq u \leq X_i)d\Lambda_{D^0}(u)$. The large sample properties of $\hat \mu(t)$ are summarized in the following Theorem.

\begin{theorem}\label{mu_thrm}
Under assumptions (A1) to (A5) in the appendix, at a fixed time point $t$, for $0\leq t \leq \tau$, 
\begin{equation*}
    n^{1/2}(\hat{\mu}(t) - \mu(t)) = n^{-1/2} \sum_{i=1}^n \Psi_i(t) + o_p(1),
\end{equation*}
where
\begin{align*}
    \Psi_i(t) = &
    \int_0^t \frac{\mu(u)dM_i^D(u)}{E[I(L\leq u \leq X)]}
    - \mu(t)\int_0^t \frac{dM_i^D(u)}{E[I(L\leq u \leq X)]} \\
    & + \int_0^t \frac{S_{D^0}(u) Y_i(u)I(L_i\leq u \leq X_i)dN_i^*(u)}{\lambda(u)} \\
    & - \int_0^t \frac{S_{D^0}(u) r(u) I(L_i\leq u \leq X_i)dN_i^*(u)}{\lambda(u)}.
\end{align*}
As $n \to \infty$, $n^{1/2}(\hat{\mu}(t) - \mu(t))$ converges in distribution to $N(0, E[\Psi_i(t)^2])$.
\end{theorem}

The estimator $\hat p(t)$ is an extension of the estimator developed by \citet{eaton_etal_2022} to left-truncated and component-wise censored data. Let $\left|\left| K_{h,t} \right|\right|_2^2 = \int_{-(\tau-u)/h}^{u/h} K_{h,u}^2(x)dx$ and $\mu_2(K_{h,u}) = \int_{-(\tau-u)/h}^{u/h} x^2K_{h,u}(x)dx$. Denote by $r^{(j)}(t)$ the $j$-th derivative of $r(t)$ with respect to $t$. Let $\lambda^{(1)}(t)$ be the first derivative of $\lambda(t)$ with respect to t. Theorem \ref{p_theorem} summarizes the large sample properties of $\hat p(t)$.

\begin{theorem}\label{p_theorem}
    Under assumptions (A1) to (A4) in the appendix, at a fixed time point $t$, for $0 \leq t \leq \tau$, $\sqrt{nh}\{\hat p(t)-p(t)-b(t,h)\}$ converges in distribution to $N(0,\sigma^2(t))$, where 
    $b(t,h) = \frac{1}{2}S_{D^0}(t) \mu_2(K_{h,t})\\\{r^{(2)}(t) + 2r^{(1)}(t) \lambda^{(1)}(t) /\lambda(t) \}h^2$, and 
    $\sigma^2(t) = S_{D^0}(t)^2 \left|\left| K_{h,t} \right|\right|_2^2 \{ r(t)-r^2(t) \} / \lambda(t)$. 
\end{theorem}

See Web Appendix A for the proofs of the asymptotic properties of the proposed estimators. To choose a proper $h$, the large sample properties of $\hat{\mu}(t)$ suggest a bandwidth on the order of $n^{-\nu}$ with $1/4 < \nu < 1/2$ (assumption (A5) in appendix). Such a bandwidth can be chosen by a data-adaptive bandwidth selection procedure as described in \citet{eaton_etal_2022} and \citet{sun_etal_2017}. The optimal bandwidth in each cross-validation has the form $h=cn^{-\nu}$. Leave-one-out cross-validation on the averaged squared error of the estimate $\hat{r}_h(t)$ and the estimand $r(t)$ is used to find $\hat{c}$. The data-adaptive selected bandwidth is then calculated by $\hat{h} = \hat{c} n^{-\nu}$.

\section{Simulation}
\label{s:sim}

\subsection{Simulation Setting} \label{s:sim_setting}

The index event onset times are generated following the simulation of \citet{wolfson_etal_2019}. We considered a constant onset rate over 10 years, and we set the prevalence day $v$ to 4 years. For the rest of section \ref{s:sim_setting}, times are defined on the personal time scale. That is, time zero represents the time of the index event. Each individual's nonfatal event time and death time are generated following the simulation of \citet{eaton_etal_2022}. The times from index event to nonfatal event, from index event to death, and from nonfatal event to death were generated from exponential distributions with rate parameters $a_{12} = 0.0008$, $a_{13} = 0.0002$, and $a_{23} = 0.0016$, respectively. All individuals are censored at the end of the study, 6 years after the prevalence day $v$. We then sample $n$ participants from the prevalent cohort, $m$ participants from the incident cohort, $n_{DO}$ from the death-only prevalent cohort, and $m_{DO}$ from the death-only incident cohort. Death-only data is created by removing all visit information $Y(t)$. 
We considered two scenarios for the visiting process. For Scenario 1 (S1), visits follow a truncated normal distribution at yearly anniversaries with a standard deviation of 30 days, truncated at $\pm$ 3 standard deviations. For Scenario 2 (S2), visits follow a stationary Poisson process with an average of one visit per year. 

First, we compared the proposed method to the KM estimator, the LT-RC-PL estimator, and the kernel estimator proposed by \citet{eaton_etal_2022} with $n=500$ prevalent cohort data. Note that in order to use the KM and LT-RC-PL estimators, we modify the component-wise censored data by defining the event time for nonfatal events as the date it was first observed at a visit. Both estimators are applied to a composite endpoint of nonfatal event and death, and if someone is not observed for the nonfatal event or death, they are censored at the end of the study. For the proposed LT-Kernel method and the kernel estimator proposed by \citet{eaton_etal_2022}, the Epanechnikov kernel with a fixed bandwidth of 1 year was used, and the boundary kernel was used in the left boundary region. A data-adaptive bandwidth selection procedure is also tested, which selects a bandwidth on the order of $n^{-7/24}$ \citep{eaton_etal_2022}. In addition, we investigated the performance of the proposed estimator under a variety of combinations of prevalent, incident, death-only prevalent, and death-only incident sample sizes $\{n,m,n_{DO},m_{DO}\}$. For example, we considered a prevalent cohort of size 500, a prevalent cohort of size 500 combined with an incident cohort of size 200, and a prevalent cohort of size 500 combined with a death-only prevalent cohort of size 200. 

All simulation results are based on 1000 simulated datasets. Per 
\citet{eaton_etal_2022}'s recommendation, the standard errors for both $\hat \mu(t)$ and $\hat p(t)$ are estimated using bootstrap (500 bootstrap samples). 

\subsection{Simulation Results} \label{s:sim_results}

\begin{figure}
\centering
\caption{\label{fig:compare_methods} Average nonfatal event-free survival (EFS) probability curves over 1000 simulated datasets with $n=500$ prevalent samples under two scenarios: S1 (panel (a)), in which potential visits occur according to a truncated normal distribution, and S2 (panel (b)), in which potential visits occur according to a stationary Poisson distribution. KM is the Kaplan-Meier estimator. LT-RC-PL is the product-limit estimator for left-truncated and right-censored data. The kernel method is the estimator proposed by Eaton et al. (2022). The LT-Kernel method is the proposed method with a fixed bandwidth of 1 year and the LT-Kernel DAB method is the proposed method with data-adaptive bandwidth selection. The curves for the LT-Kernel estimators are almost identical to one another and to the true EFS curve.}
\includegraphics[width=\linewidth]{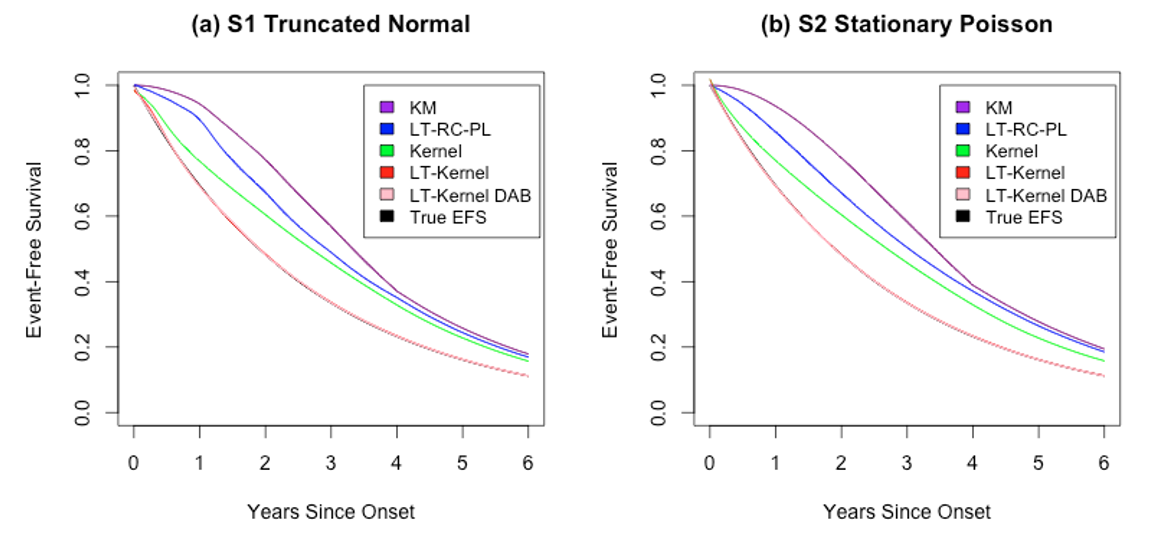}
\end{figure}

Under both scenarios, the LT-Kernel estimators had the smallest bias and mean squared error for both $p(3)$ and $\mu(3)$ (Table \ref{tab:compare_methods_mu}). Using a fixed bandwidth or data-adaptive bandwidth performed similarly, with the average selected bandwidth being 0.52 years for S1 and 0.55 years for S2. The average estimated EFS curves from the LT-Kernel estimator with fixed or data-adaptive bandwidth were closest to the truth (Figure \ref{fig:compare_methods}). Figure 2 also showed that the average estimated EFS curve of the kernel estimator proposed by \citet{eaton_etal_2022} had smaller bias compared to the curve of LT-RC-PL, suggesting that failing to account for component-wise censoring has a greater impact than left-truncation in this simulation setting. The variance estimates were accurate, with the empirical standard error (SE) generally consistent with the mean standard error estimate averaged over simulated datasets (SEE). Results showing the performance of the estimators at year 5 were similar and can be found in Web Table 1 (Web Appendix B).

\begin{table}[]
    \centering
    \caption{Performance of estimators of nonfatal event-free survival probability and EF-RMST  at 3 years post index event, with 1000 simulated datasets with $n=500$ prevalent sample. The performance was evaluated under two scenarios: S1 (top panel), in which potential visits occur according to a truncated normal distribution, and S2 (bottom panel), in which potential visits occur according to a stationary Poisson distribution. True values of $p(3)$ and $\mu(3)$ are shown in the table's header. }
\resizebox{0.9\linewidth}{!}{
\begin{threeparttable}
    \begin{tabular}{l|rrrr|rrrr}
    \hline
    \hline
    \multicolumn{1}{l|}{S1 Truncated Normal} & \multicolumn{4}{c|}{$p(3) = 0.3345$} & \multicolumn{4}{c}{$\mu(3) = 1.8232$ years} \\
    \hline
    Methods & Bias & SE & SEE & MSE & Bias & SE & SEE & MSE\\
    \hline
    KM                  & 0.2342 & 0.0214 & 0.0221 & 0.0553 &
                          0.6876 & 0.0314 & 0.0326 & 0.4738 \\
    LT-RC-PL            & 0.1548 & 0.0344 & 0.0342 & 0.0251 &
                          0.4828 & 0.1002 & 0.0881 & 0.2431 \\
    Kernel Method       & 0.1227 & 0.0211 & 0.0220 & 0.0155 & 
                          0.2685 & 0.0539 & 0.0554 & 0.0750 \\
    LT-Kernel Method    & 0.0023 & 0.0251 & 0.0256 & 0.0006 &
                          0.0047 & 0.0923 & 0.0872 & 0.0085 \\
    LT-Kernel DAB       & 0.0006 & 0.0267 & 0.0268 & 0.0007 &
                          0.0028 & 0.0918 & 0.0875 & 0.0084 \\ 
    \hline
    \hline
    \multicolumn{1}{l|}{S2 Stationary Poisson} & \multicolumn{4}{c|}{$p(3) = 0.3345$} & \multicolumn{4}{c}{$\mu(3) = 1.8232$ years} \\
    \hline
    Methods & Bias & SE & SEE & MSE & Bias & SE & SEE & MSE\\ 
    \hline
    KM                  & 0.2482 & 0.0216 & 0.0220 & 0.0621 &
                          0.6988 & 0.0328 & 0.0329 & 0.4894 \\
    LT-RC-PL            & 0.1696 & 0.0387 & 0.0372 & 0.0303 &
                          0.4631 & 0.1262 & 0.1037 & 0.2304 \\
    Kernel Method       & 0.1218 & 0.0287 & 0.0278 & 0.0157 &
                          0.2805 & 0.1981 & 0.1504 & 0.1179 \\
    LT-Kernel Method    & 0.0015 & 0.0284 & 0.0277 & 0.0008 &
                          0.0001 & 0.1070 & 0.0929 & 0.0114 \\
    LT-Kernel DAB       & 0.0001 & 0.0310 & 0.0298 & 0.0010 &
                         -0.0033 & 0.1063 & 0.0934 & 0.0113 \\
    \hline
    \hline
    \end{tabular}
    \begin{tablenotes}
        \setlength\labelsep{0pt}
        \small
        \item Table abbreviations: KM is the Kaplan-Meier estimator, and LT-RC-PL is the product-limit estimator for left-truncated and right-censored data. The kernel method is the estimator proposed by Eaton et al. (2022). The LT-Kernel method is the proposed method with a fixed bandwidth of 1 year. LT-Kernel DAB is the proposed method with a data-adaptive bandwidth. SE is the empirical standard error. SEE is the mean of the standard error estimates. MSE is the mean squared error. 
    \end{tablenotes}
    \end{threeparttable}
}
    \label{tab:compare_methods_mu}
\end{table}

\begin{figure}
\centering
\caption{\label{fig:compare_data} Average empirical standard error (SE) estimation of the nonfatal event-free survival probabilities with 1000 simulated datasets under S1, in which potential visits occur according to a truncated normal distribution. $200$ (light blue) or $500$ (blue) samples were added in addition to the given $n=500$ prevalent samples from different data sources, shown in each panel: (a) for incident data; (b) for death-only incident data; (c) for death-only prevalent data. }
\includegraphics[width=1\linewidth]{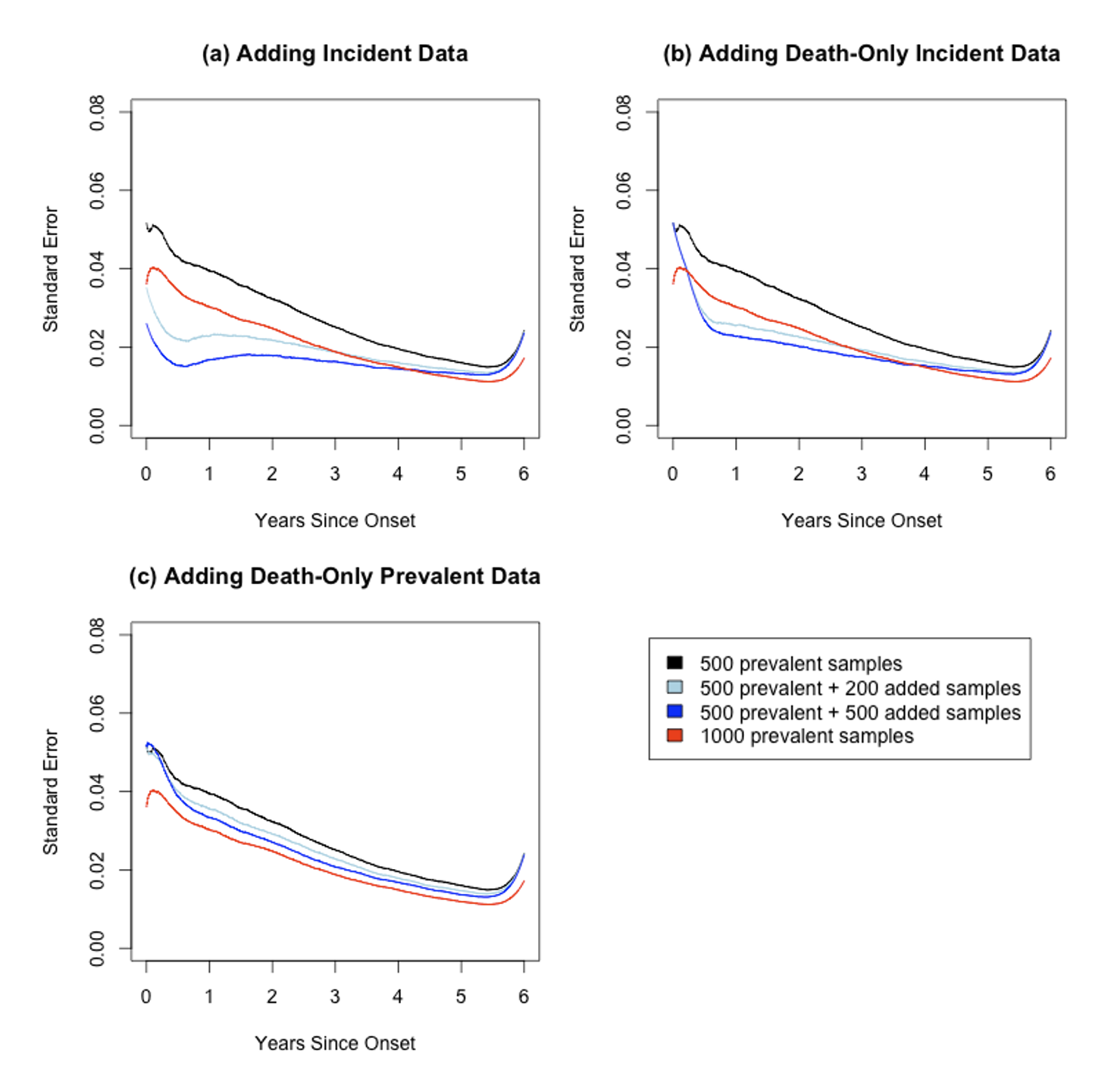}
\end{figure}

We then investigated how incorporating additional data sources affects EFS estimation, as shown in Table \ref{tab:compare_data} and Figure \ref{fig:compare_data}. Under S1 and using a fixed bandwidth of 1-year for the LT-Kernel method, the estimates of $p(3)$ and $p(5)$ remained close to the truth with small biases across all combinations of data types and sample sizes. We observed the largest gain in precision by adding incident data. Specifically, when estimating $p(3)$, adding only 200 incident data to 500 prevalent data ($n=500, m=200$) achieved a standard error comparable to doubling the prevalent cohort to 1000 ($n=1000$). Adding incident data reduced the variance of the estimator considerably for earlier follow-up times, when the risk sets from the prevalent cohort data tend to be small (Figure \ref{fig:compare_data}, panel (a)).

We observed a similar but smaller reduction in variance by adding death-only incident data (Table \ref{tab:compare_data}). By adding 500 death-only incident cases ($n=500, m_{DO}=500$),  the estimator performed comparably to doubling the prevalent sample alone ($n=1000$), and again, early-year variance was reduced. Because death-only data contain less information than complete incident data, the variance reduction was smaller (Figure \ref{fig:compare_data}, comparing panel (b) to (a)).

Finally, adding prevalent death-only data had a smaller impact on the variance of the estimator. The two SE curves with additional death-only prevalent data ($n=500, n_{DO}=200$ and $n=500, n_{DO}=500$) fall between the curves from the $n=500$ and $n=1000$ prevalent settings. The results are as expected, as prevalent death-only cases contain less information than complete prevalent cases (with data on both death and the non-fatal event). Adding 500 prevalent death-only samples provides additional information above that provided by $n=500$ prevalent samples alone, but not as much information as $n=1000$ prevalent samples. (Table \ref{tab:compare_data}, Figure \ref{fig:compare_data}, panel (c)). 

\begin{table}[]
    \centering
    \caption{Effect of sample size and data source on estimating nonfatal event-free survival probability at 3 and 5 years post index event with the LT-Kernel estimator with a fixed bandwidth at 1-year, with 1000 simulated datasets. True values of $p(3)$ and $p(5)$ are shown in the table's header. The first row shows results using $n=500$ prevalent samples;  in subsequent rows, $200$ to $500$ samples were added from different data sources; $m$ represents additional incident cases, $n_{DO}$ represents additional death-only prevalent cases, and $m_{DO}$ represents additional death-only incident cases. }
\resizebox{\linewidth}{!}{
\begin{threeparttable}
    \begin{tabular}{l|rrrr|rrrr}
    \hline
    \hline
    \multicolumn{1}{c|}{} & \multicolumn{4}{c|}{$p(3)=0.3345$} & \multicolumn{4}{c}{$p(5)=0.1612$} \\
    \hline
    Data Source and Sample Size & Bias & SE & SEE & MSE& Bias & SE & SEE & MSE\\
    \hline
    \multicolumn{1}{l|}{Prevalent Only} & \multicolumn{4}{l|}{} & \multicolumn{4}{l}{}\\
    $n=500$             & 0.0023 & 0.0251 & 0.0257 & 0.0006
                        & 0.0012 & 0.0160 & 0.0164 & 0.0003 \\
    $n=1000$            & 0.0021 & 0.0188 & 0.0185 & 0.0004 
                        & 0.0015 & 0.0119 & 0.0117 & 0.0001 \\
    \hline
    \multicolumn{1}{l|}{Adding Incident Data} & \multicolumn{4}{l|}{} & \multicolumn{4}{l}{} \\
    $n=500,m=200$       & 0.0021 & 0.0188 & 0.0195 & 0.0004
                        & 0.0011 & 0.0140 & 0.0141 & 0.0002 \\
    $n=500,m=500$       & 0.0027 & 0.0163 & 0.0166 & 0.0003 
                        & 0.0012 & 0.0132 & 0.0129 & 0.0002 \\
    \hline
    \multicolumn{1}{l|}{Adding Death-Only Incident Data}  & \multicolumn{4}{l|}{} & \multicolumn{4}{l}{}\\
    $n=500, m_{DO}=200$ & 0.0017 & 0.0193 & 0.0201 & 0.0004
                        & 0.0010 & 0.0141 & 0.0142 & 0.0002 \\
    $n=500, m_{DO}=500$ & 0.0018 & 0.0175 & 0.0181 & 0.0003 
                        & 0.0011 & 0.0136 & 0.0134 & 0.0002 \\
    \hline
    \multicolumn{1}{l|}{Adding Death-Only Prevalent Data}  & \multicolumn{4}{l|}{} & \multicolumn{4}{l}{}\\
    $n=500, n_{DO}=200$ & 0.0022 & 0.0228 & 0.0229 & 0.0005
                        & 0.0014 & 0.0147 & 0.0148 & 0.0002 \\
    $n=500, n_{DO}=500$ & 0.0020 & 0.0207 & 0.0207 & 0.0004 
                        & 0.0014 & 0.0137 & 0.0136 & 0.0002 \\

    \hline
    \hline
    \end{tabular}
    \begin{tablenotes}
        \setlength\labelsep{0pt}
        \small
        \item Table abbreviations: SE is the empirical standard error of the nonfatal event-free survival probability. SEE is the mean of the standard error estimates. MSE is the mean squared error. 
    \end{tablenotes}
    \end{threeparttable}
}
\label{tab:compare_data}
\end{table}

\section{Real Data Analysis}
\label{s:rda}

The ARIC study is a community-based surveillance study focusing on atherosclerotic cardiovascular disease. The study was launched in 1987 (Visit 1, 1987 to 1989) and recruited 15,792 predominantly Black and white middle-aged adults from four US communities (Forsyth County, NC, Jackson, MS, suburban Minneapolis, MN, and Washington County, MD). Over time, additional endpoints and risk factors were added to the study, including dementia assessment as the study population ages \citep{wright_etal_2021}. 
We are interested in estimating the post-MI dementia-free survival (DemFS) probability, and short-term (5-year) and long-term (20-year) restricted mean dementia-free survival time (DemF-RMST) post-MI among the ARIC population. Further, we would like to test for differences in DemF-RMST between subgroups of participants, including differences by race, sex, and age.

MI was defined as acute and hospitalized definite or probable MI adjudicated by ARIC using standardized algorithms based on chest pain symptoms, ECG findings, and cardiac biomarker measurements, following the definition in \citet{myerson_etal_2009}. Since we are interested in post-MI survival outcomes for those who survive their initial MI, we excluded fatal MIs and participants who passed away within 3 days of their MI. Visit 5 (2011 to 2013) was chosen as the baseline (i.e., prevalence day) for this data analysis because, starting at Visit 5, ARIC started to implement neuropsychological cognitive testing, the Atherosclerosis Risk in Communities Neuro-cognitive Study (ARIC-NCS), at visits \citep{knopman_etal_2016}. 
We chose to use the Level-1 dementia diagnosis, which was based on neuropsychological tests at scheduled ARIC study visits in-person (with the exception of phone-based neuropsychological tests at Visit 8, 2020).  Dementia status was thus based on scheduled assessments from Visit 5 to Visit 9 (2021 to 2022). We chose not to use dementia diagnoses recorded during hospitalizations or on death certificates, because treating these as a ``visit" would likely lead to dependence between the true time of dementia onset and the visit schedule, as people may be more likely to be hospitalized or die following the onset of dementia. This would violate the conditional independent visit process assumption of the proposed method.

Participants needed to be alive at Visit 5 to be included in this analysis. With data up to Visit 9, 1336 participants had an MI recorded. 811 participants had their MI before ARIC Visit 5 (prevalent cohort), including 453 who attended at least one ARIC visit after their MI and 359 who did not. 525 participants had their MI after Visit 5 (incident cohort), including 166 who attended at least one ARIC visit after their MI and 359 who did not. A demonstrative figure for 10 ARIC participants (7 from the prevalent cohort and 3 from the incident cohort) can be found in Web Figure 1 (Web Appendix B).

We applied the proposed method to the ARIC cohort with all 1336 participants described above to estimate the DemFS probability and DemF-RMST at 5 and 20 years post-MI using the Epanechnikov kernel function, with a fixed bandwidth of 3 years based on the study design and a data-adaptive bandwidth selected at 5.8 years. Although the theoretical longest follow-up time is 36 years, we chose $\tau=30$ years, earlier than the largest failure time observed, to ensure the DemF-RMST is estimable. Bootstrap percentile confidence intervals (CI) based on 500 bootstrap samples were used. Using the proposed method with the fixed 3-year bandwidth, we estimated that a participant is expected to live dementia-free 2.87 (95\% CI: 2.68 to 3.06) years out of the next 5 years after their first MI on average, or 5.72 (95\% CI: 5.27 to 6.14) years out of the next 20 years after their first MI on average. Using a data-adaptively selected bandwidth of 5.8 years provided very similar results (Web Appendix B, Web Table 2).

\begin{figure}
\centering
\caption{\label{fig:RDA_wa} 
Estimated dementia-free survival probability up to 25 years post-MI using the proposed methods and the comparison methods used in simulation studies in the ARIC population. LT-RC-PL is the product-limit estimator for left-truncated and right-censored data. The kernel method is the estimator proposed by Eaton et al. (2022). The proposed LT-Kernel method is shown with a fixed bandwidth $h=3$ years, or a data-adaptively selected bandwidth (DAB) of 5.8 years.}
\includegraphics[width=0.7\linewidth]{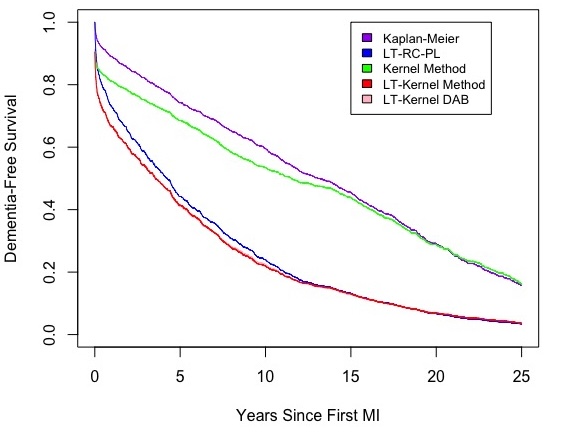}
\end{figure}

We then compared the results of the proposed method to the three comparison methods introduced in Section \ref{s:sim_setting}. For the KM estimator and the LT-RC-PL estimator, the dementia date was assigned to the visit date at which dementia is confirmed. Both estimators are applied to a composite endpoint of dementia and death, and participants who are not observed to experience either event are censored at the last contact or the end of follow-up at Visit 9. We used the same Epanechnikov kernel function and a fixed 3-year bandwidth for the kernel estimator proposed by \citet{eaton_etal_2022}. Figure \ref{fig:RDA_wa} shows estimated DemFS up to 25 years post-MI. All three comparison methods yielded larger estimates for the post-MI DemFS, and the proposed method with either a fixed 3-year bandwidth or a data-adaptively selected 5.8-year bandwidth showed almost identical results (Figure \ref{fig:RDA_wa}), similar to the simulation study. Different from the simulation results, the kernel method proposed by \citet{eaton_etal_2022} yielded a larger estimate than the LT-RC-PL estimator, suggesting that left-truncation, if not properly taken care of, is a more severe problem than interval censoring in the ARIC data. This result is as expected, as the left truncation time of the ARIC data is longer relative to the dementia-free survival time scale, compared to the simulation study. Specifically, for the ARIC data, the longest observed left truncation time is 24.7 years, compared to a median dementia-free survival time of 3.5 years (estimated by the proposed method with the fixed bandwidth), out of a maximum observed follow-up time of 35.2 years. Meanwhile, in simulation studies, the longest left truncation time is 4 years, compared to a median event-free survival time of 1.9 years, out of a maximum follow-up time of 10 years by design. A more detailed table consisting of the estimated DemFS and DemF-RMST and their variance at 5 and 20 years post-MI using the proposed and comparison methods can be found in Web Table 2 (Web Appendix B). 

Finally, we test for differences in DemF-RMST using the proposed estimator, testing the null hypothesis that the DemF-RMST are equal in the short- (5-year) and long-term (20-year) between sexes, racial groups, and age groups dichotomized at the median age of MI onset  ($<$ or $\geq$ 74) and report the z-test p-values. The DemF-RMST of male participants and female participants is not significantly different at either time point. Male participants were expected to live 0.18 years less in the next 5 years post-MI (95\% CI: -0.55, 0.24; $p=0.36$) and 0.80 years less in the next 20 years post-MI (95\% CI: -1.72, 0.18; $p=0.092$), compared to female participants. The DemF-RMST of Black participants and white participants is also not significantly different at either time point. Black participants were expected to live 0.29 years less (95\% CI: -0.86, 0.21) and 1.07 years less (95\% CI: -2.02, 0.25) in the next 5 and 20 years post-MI, with a p-value of 0.28 and 0.11, respectively, compared to white participants. The DemF-RMST of younger participants (age at MI $<74$) and older patients (age at MI $\geq 74$) is not significantly different in the short-term. Younger participants are expected to live 0.50 years longer in the next 5 years post-MI (95\% CI: -0.01, 1.09; p=0.078) compared to older participants. However, we observed significant DemF-RMST differences in the long term. Younger participants are expected to live 3.51 years longer in the next 20 years post-MI (95\% CI: 2.12, 5.02; $p<0.0001$) compared to older participants. A more detailed summary of the estimated DemF-RMST of each subgroup and the differences between the two groups can be found in Web Table 3 (Web Appendix B).

\section{Discussion}
\label{s:discuss}

Component-wise censoring arises when a composite endpoint comprising multiple time-to-event endpoints is of interest, when some component(s) of the endpoints can only be measured intermittently. This setting could arise in different designs, including prospective clinical trials or cohort studies, as well as retrospectively collected real-world data such as electronic health records or registry data. 

It is common in practice to convert component-wise censoring to right-censoring when analyzing treatment effects, but \citet{carroll_2007} showed that this approach leads to bias in treatment effects depending on the visit schedules and the expected EFS time. In the setting where two treatment groups follow the same visit schedule, although assigning event times to the first detection visit and analyzing them using standard right-censored methods yields valid log-rank tests when the required number of events is increased to preserve power \citep{carroll_2007}, the resulting group-specific estimates are still biased due to interval censoring. Our proposed method addressed this limitation by accounting for left truncation and component-wise censoring through separating the estimand into two components, one related to the interval-censored nonfatal event and one related to the left-truncated and right-censored death event. The proposed method is robust and flexible as it does not require any assumption about the dependence between the interval-censored and right-censored endpoints, and is not limited to composite endpoints consisting of exactly one interval-censored nonfatal event and one right-censored death. For example, in the real data analysis presented by \citet{eaton_etal_2022}, the composite endpoint had an interval-censored component of multiple nonfatal events that were followed up annually, and a right-censored component of death and a nonfatal event that was followed continuously and thus also right-censored. Our proposed method can be formatted the same way and provide accurate estimates of all-nonfatal-event-free survival probability or restricted mean all-nonfatal-event-free survival time. In addition, \citet{eaton_etal_2022} discuss using their kernel estimator (for data without left truncation) to estimate the probability in each state of a multistate model, including models with reversible transitions, when the current state can only be observed intermittently. Our proposed estimator could be extended in the same way to handle more complex endpoints. 

Multiple studies have examined the benefits of using period prevalent data (i.e., mixed incident and prevalent cohort data). \citet{wolfson_etal_2019} explored the accuracy gain afforded by adding incident data to prevalent data when estimating survival probability, and \citet{hartman_2025} found the optimal mix for the most accurate survival probability curve. The Nun Study explored the improved stability in risk factor identification under the proportion mean residual life model \citep{lee_etal_2019}. To our knowledge, the benefits of supplementing data from a prevalent cohort study with death-only data have not been studied. While there is no obvious way to integrate death-only data into the estimation of EFS using the standard right-censored approach, because our estimator has one part related to overall survival only, it can naturally leverage death-only data. In line with the existing literature, our proposed estimator was minimally biased for all combinations of samples. When the prevalent cohort data were supplemented with either complete incident cohort data (which included visits to assess the nonfatal event) or death-only incident data (with no visits by design), we observed reductions in the variance of the estimator, especially the best improvement in variance estimation with supplemental incident cohort data. The good performance of our proposed estimator shows a promising future in utilizing the vast amount of existing real-world data for biomedical research. In addition to the statistical benefits, combining data sources enables researchers to capture unique patients from a larger range of populations under routine medical care \citep{levenson_etal_2023}.  

While combining data sources can be a powerful approach, care must be taken to ensure that the datasets being combined represent the same underlying population. In addition, real-world datasets present a risk of violating the independent visit assumption needed for the proposed estimator. Unlike in clinical trials, where the visit schedules are planned ahead, in real-world data, especially in electronic health record data, a visit can be event-driven (e.g., a patient schedules a nonfatal event assessment because they are experiencing symptoms of the event). A patient’s visit rate or their probability of exiting the study (right-censoring) could also depend on patient covariates; if these covariates are also related to death or the nonfatal event, this would violate the assumptions of our estimator. 

Another common limitation of using real-world data is that a set of desired baseline covariate values may not be available. This is in contrast to clinical trials, where participants are enrolled in the study and all covariates of interest are measured at their baseline visit. An implication of this in the ARIC data analysis was that we were unable to compare DemF-RMST between groups based on time-varying disease-related biomarkers or modifiable risk factors; instead, we compared groups defined by stable participant characteristics.

Future work includes examining the robustness of the estimators to assumption violations and extending the proposed estimators to remain valid under relaxed independence assumptions. In addition, as \citet{wolfson_etal_2019} mentioned and \citet{hartman_2025} studied, an optimal proportion of prevalent, incident, and death-only cohorts can be developed for sample size allocation to achieve the most efficient study design with left-truncated death and component-wise censored endpoint. Moreover, while the proposed estimator provided a z-test for differences of EF-RMST in subgroups, we would like to understand the effect of a continuous variable (e.g., age as a continuous variable, not dichotomized in ARIC) or the joint effect of multiple variables on EFS or EF-RMST. Developing semiparametric regression models for the left-truncated and component-wise censored data shall also be considered as a future direction.

\section*{Acknowledgements}

The Atherosclerosis Risk in Communities Study is carried out as a collaborative study supported by National Heart, Lung, and Blood Institute contracts (75N92022D00001, 75N92022D00002, 75N92022D00003, 75N92022D00004, 75N92022D00005). The ARIC Neurocognitive Study is supported by U01HL096812, U01HL096814, U01HL096899, U01HL096902, and U01HL096917 from the NIH (NHLBI, NINDS, NIA and NIDCD). The authors thank the staff and participants of the ARIC study for their important contributions. Data that support the findings of this study are available in accordance with ARIC study policies. Data requests can be assessed at \url{https://aric.cscc.unc.edu/aric9/researchers/Obtain_Submit_Data} for ARIC data. The authors acknowledge the Minnesota Supercomputing Institute (MSI) at the University of Minnesota for providing resources that contributed to the research results reported within this paper. 

\section*{Supporting information}
Web Appendix A and B, referenced in Section \ref{s:lsp}, \ref{s:sim_results}, and \ref{s:rda}, are available with this paper at the end of the document. The R package \textbf{ltkernelcwc} for the proposed method is available on GitHub at \url{https://github.com/HanLu-98/ltkernelcwc}.

\newpage

\setcounter{section}{0}
\setcounter{table}{0}
\renewcommand{\tablename}{Web Table}
\setcounter{figure}{0}
\renewcommand{\figurename}{Web Figure}

\section*{Appendix}
\section{Regularity Conditions}

Assumptions (A1) to (A5) are the regularity conditions needed for the large sample properties

(A1) The censoring time $C_i$ and the left truncation time $L_i$ are independent of $\{D_i, N_i^*(\cdot), \\Y_i(\cdot)\}$ and $\inf_{t\in[0,\tau]}P(L_i \leq t\leq X_i)>0$. 

(A2) The counting process $N_i^*(t)$ is independent of $\{L_i,D_i,C_i,Y_i(\cdot)\}$. The observation time process $I(L_i\leq t \leq X_i)$ is bounded, and the second derivative of its rate function $\lambda(t)$ is bounded. Moreover, $\lambda(t)>0$ for $t\in[0,\tau]$.

(A3) Define $\xi(t)$ such that $\xi(t)dt = E[Y(t)I(L\leq t\leq X)dN^*(t)]$, and the second derivative of $\xi(t)$ is bounded for $t\in[0,\tau]$.

(A4) For $u \in [0,\tau]$, the kernel $K_{h,u}(\cdot)$ satisfies $\int_{-(\tau-u)/h}^{u/h} K_{h,u}(x)dx = 1$ and $\int_{-(\tau-u)/h}^{u/h} xK_{h,u}(x)dx = 0$. 

(A5) The bandwidth h satisfies the following property: $h\asymp n^{-\nu},1/4<\nu < 1/2$.

\section{Web Appendix A: Proof of Theorem 1 and Theorem 2}

\subsection{Proof for Theorem 1}

First, we showed that the population survival mean of the nonfatal event is the same as the sample survival mean, i.e., $E[Y^0(t)|D^0\geq t] = E[Y(t)|dN(t)=1]$, under independent left truncation, conditional independent residual censoring, and conditional independent visit process assumptions mentioned in the Large Sample Properties section of the manuscript. 
\begin{align*}
    &P\{ Y(t) \leq y | dN(t) = 1 \} \\
    =& P\{ Y(t) \leq y | dN^*(t-L)=1, D \geq t, C_{res} \geq t-L, L \leq t \} \\
    =& \frac{P\{ Y(t) \leq y, dN^*(t-L)=1, D \geq t, C_{res} \geq t-L , L \leq t \}}
    {P\{ dN^*(t-L)=1, D \geq t, C_{res} \geq t-L , L \leq t \}} \\
    =& \frac{\int_0^t P\{ Y(t) \leq y, dN^*(t-l)=1, D\geq t, C_{res} \geq t-l | L=l \} f_L(l)dl}
    {\int_0^t P\{ dN^*(t-l)=1, D\geq t, C_{res} \geq t-l | L=l \} f_L(l)dl} \\
    =& \frac{P\{ Y^0(t) \leq y, D^0\geq t| L^0=l \} \int_0^t P\{  dN^*(t-l)=1,C_{res} \geq t-l |L=l \}f_L(l)dl}
    {P\{D^0\geq t| L^0=l \}\int_0^t  P\{ dN^*(t-l)=1,C_{res} \geq t-l |L=l\}f_L(l)dl} \\
    =& \frac{P\{Y^0(t) \leq y, D^0\geq t\}}{P(D^0\geq t)} = P\{Y^0(t) \leq y| D^0\geq t\} \\
    & \text{Therefore, }E[Y^0(t)|D^0\geq t] \\
    =& \sum_{y\in\{0,1\}} 1-P\{ Y^0(t) \leq y | D^0(t) \geq t \} \\
    =& \sum_{y\in\{0,1\}} 1-P\{Y(t) \leq y | dN(t)=1 \} \\
    =& E[Y(t)|dN(t)=1]
\end{align*}

Then, the proof of Theorem 1 follows the proof of Theorem 3.1 in Sun et al. (2017) with modifications. To accommodate data with left truncation, we modify the indicator functions from $I(X_i \geq u)$ to $I(L_i \leq u \leq X_i)$ in Lemmas 1, 2, 3, and in the proof. We further specify $S_X(u) = E[I(L \leq u \leq X)]$, $\hat S_X(u) = \frac{1}{n}\sum_{i=1}^n I(L_i \leq u \leq X_i)$ and $w(u)=1$. In addition, $\hat{S}_{D_0}$ has a similar independent and identically distributed (i.i.d.) representation with left-truncated and right-censored data (Gijbels and Wang, 1993) to the Kaplan-Meier estimator for right-censored data used in Sun et al. (2017). Specifically, under our assumptions, $\sqrt{n}(\hat{S}_{D^0}(t) - S_{D^0}(t)) = n^{-1/2}\sum_{i=1}^n \phi_i^D(t) + o_p(1)$ where $\phi_i^D(t) = -S_{D^0}(t)\int_0^t dM_i^D(u)/E[I(L\leq u\leq X)]$. With the above modifications and the asymptotic property unchanged, the same proof follows.

\subsection{Proof for Theorem 2}

The proof of Theorem 2 follows the proof of Theorem 2 in Eaton et al. (2022) with modifications. The definition of $N_i(t)$ now incorporates left truncation time, defined as $N_i(t) = I(L_i \leq t \leq X_i)dN_i^*(t)$. Again, the i.i.d. representation holds for the product-limit estimator for left-truncated and right-censored data (Gijbels and Wang, 1993). With the modified definition of the observed counting process and the asymptotic results, the same proof follows.

\section{Web Appendix B: Supplemental Methods and Results}

\subsection{Simulation results for the proposed estimators at year-5}

Web Table \ref{tab:compare_methods_mu_wa_yr5} shows the performance of the estimators at year 5 under both scenarios.

\begin{table}[]
    \centering
    \caption{Performance of estimators of nonfatal event-free survival probability and EF-RMST  at 5 years post index event, with 1000 simulated datasets with $n=500$ prevalent sample. The performance was evaluated under two scenarios: S1 (top panel), in which potential visits occur according to a truncated normal distribution, and S2 (bottom panel), in which potential visits occur according to a stationary Poisson distribution. True values of $p(5)$ and $\mu(5)$ are shown in the table's header.}
\resizebox{\linewidth}{!}{
\begin{threeparttable}
    \begin{tabular}{l|rrrr|rrrr}
    \hline
    \hline
    \multicolumn{1}{l|}{S1 Truncated Normal} & \multicolumn{4}{c|}{$p(5) = 0.1612$} & \multicolumn{4}{c}{$\mu(5) = 2.2980$ years} \\
    \hline
    Methods & Bias & SE & SEE & MSE & Bias & SE & SEE & MSE\\
    \hline
    KM                  & 0.0965 & 0.0191 & 0.0195 & 0.0097 & 
                          0.9906 & 0.0605 & 0.0638 & 0.9850 \\
    LT-RC-PL            & 0.0829 & 0.0227 & 0.0230 & 0.0074 &
                          0.7177 & 0.1447 & 0.1345 & 0.5361 \\
    Kernel Method       & 0.0666 & 0.0182 & 0.0184 & 0.0048 &
                          0.4602 & 0.0785 & 0.0824 & 0.2179 \\
    LT-Kernel Method    & 0.0012 & 0.0160 & 0.0164 & 0.0003 &
                          0.0085 & 0.1233 & 0.1199 & 0.0153 \\
    LT-Kernel DAB       & 0.0007 & 0.0171 & 0.0172 & 0.0003 &
                          0.0049 & 0.1226 & 0.1200 & 0.0150 \\
    \hline
    \hline
    \multicolumn{1}{l|}{S2 Stationary Poisson} & \multicolumn{4}{c|}{$p(5) = 0.1612$} & \multicolumn{4}{c}{$\mu(5) = 2.2980$ years} \\
    \hline
    Methods & Bias & SE & SEE & MSE & Bias & SE & SEE & MSE\\
    \hline
    KM                  & 0.1165 & 0.0196 & 0.0200 & 0.0139 &
                          1.0397 & 0.0640 & 0.0644 & 1.0850 \\
    LT-RC-PL            & 0.1034 & 0.0253 & 0.0250 & 0.0113 &
                          0.7385 & 0.1800 & 0.1556 & 0.5777 \\
    Kernel Method       & 0.0668 & 0.0212 & 0.0212 & 0.0049 &
                          0.4719 & 0.2112 & 0.1677 & 0.2673 \\
    LT-Kernel Method    & 0.0006 & 0.0182 & 0.0183 & 0.0003 &
                          0.0030 & 0.1422 & 0.1268 & 0.0202 \\
    LT-Kernel DAB       & 0.0001 & 0.0200 & 0.0200 & 0.0004 &
                         -0.0024 & 0.1414 & 0.1270 & 0.0200 \\
    \hline
    \hline
    \end{tabular}
    \begin{tablenotes}
        \setlength\labelsep{0pt}
        \small
        \item Table abbreviations: KM is the Kaplan-Meier estimator, and LT-RC-PL is the product-limit estimator for left-truncated and right-censored data. The kernel method is the estimator proposed by Eaton et al. (2022). The LT-Kernel method is the proposed method with a fixed bandwidth of 1 year. LT-Kernel DAB is the proposed method with a data-adaptive bandwidth. SE is the empirical standard error. SEE is the mean of the standard error estimates. MSE is the mean squared error.
    \end{tablenotes}
    \end{threeparttable}
}
    \label{tab:compare_methods_mu_wa_yr5}
\end{table}

\subsection{Swimmer's plot for selected ARIC participants}
\begin{figure}
\centering
\caption{\label{fig:RDA_demo} 
Swimmer plot of 10 selected ARIC participants in prevalent and incident cohorts under the calendar time scale (left panel) and the personal time scale (right panel). On a calendar scale, time $0$ represents the earliest possible Visit 1 date, January 1, 1987. }
\includegraphics[width=\linewidth]{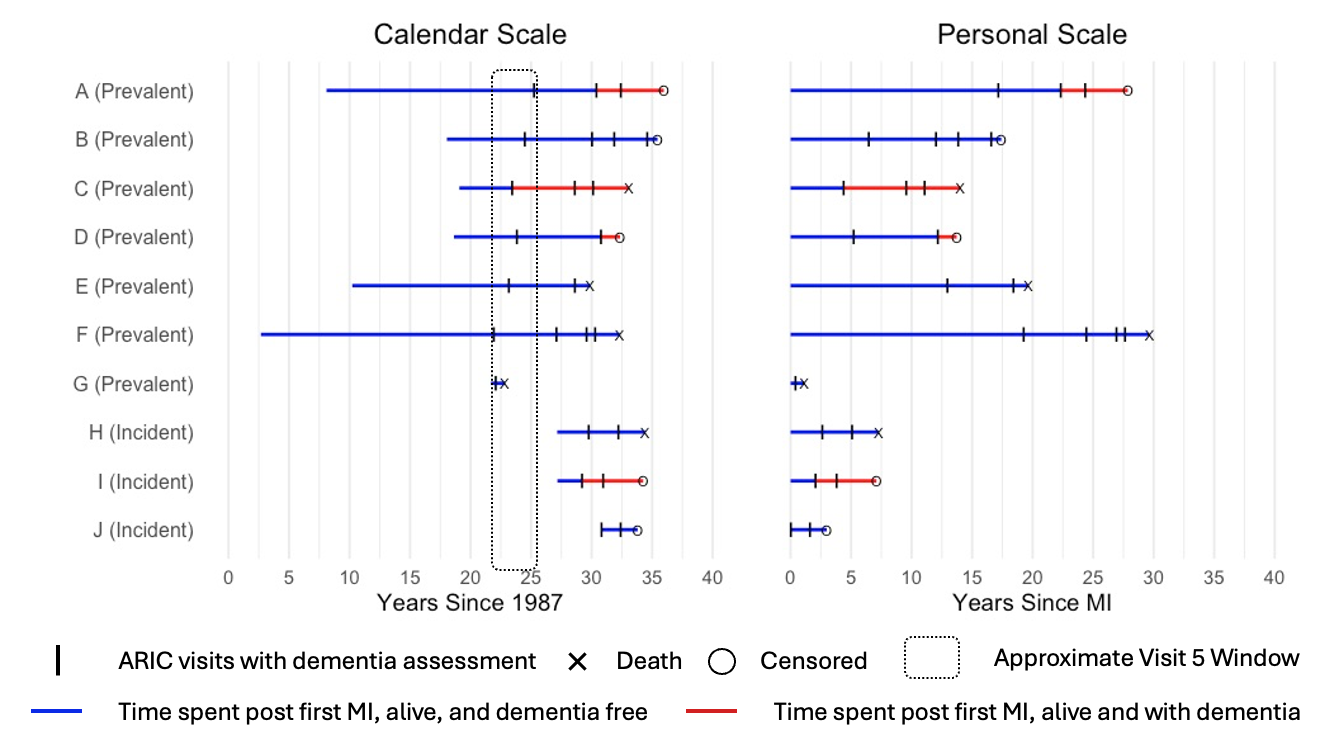}
\end{figure}

We plotted the first MI, ARIC visits, dementia, and death or censoring time for 10 ARIC participants, including 7 participants from the prevalent cohort who attended at least one ARIC visit on or after Visit 5, and 3 participants from the incident cohort who attended at least one ARIC visit on or after Visit 5 (Web Figure \ref{fig:RDA_demo}). 
The available ARIC data did not include exact event or visit dates; instead, the number of days from a participant’s first ARIC visit was recorded. To convert ARIC data to the calendar time scale, we assumed that each person’s first visit occurred on January 1, 1987. In practice, first visits occurred between 1987 and 1989.   

\subsection{Estimated DemFS and DemF-RMST in the ARIC data with additional comparison methods}

We estimated DemFS and DemF-RMST at 5 and 20 years post-MI with the ARIC data, with the results shown in Web Table \ref{tab:rda_table_wa}.

\begin{table}[]
\centering
\caption{Estimated dementia-free survival probability and restricted mean dementia-free survival time at 5 and 20 years post-MI using the comparison methods, the proposed LT-Kernel method with a fixed bandwidth $h=3$ years, and the proposed LT-Kernel method with data-adaptive bandwidth selection (DAB) with a selected bandwidth of 5.8 years. }
\resizebox{\linewidth}{!}{
\begin{threeparttable}
\begin{tabular}{l|rrrr|rrrr}
  \hline
  \hline
 & $\hat{p}(5)$ & SE & $\hat{\mu}(5)$ & SE & $\hat{p}(20)$ & SE & $\hat{\mu}(20)$ & SE  \\ 
  \hline
KM & 0.7426 & 0.0120 & 4.1874 & 0.0428 & 0.2910 & 0.0139 & 11.9779 & 0.2057 \\ 
LT-RC-PL & 0.4421 & 0.0201 & 3.1211 & 0.0850 & 0.0673 & 0.0059 & 6.1436 & 0.2247 \\ 
Kernel Method & 0.6851 & 0.0176 & 3.8315 & 0.0814 & 0.2876 & 0.0148 & 11.0895 & 0.2438 \\ 
LT-Kernel Method & 0.4138 & 0.0204 & 2.8750 & 0.0975 & 0.0690 & 0.0060 & 5.7283 & 0.2329 \\ 
LT-Kernel Method DAB & 0.4115 & 0.0196 & 2.8735 & 0.0946 & 0.0701 & 0.0060 & 5.7335 & 0.2323 \\ 
   \hline
\end{tabular}
\begin{tablenotes}
        \setlength\labelsep{0pt}
        \small
        \item Table abbreviations: KM is the Kaplan-Meier estimator, and LT-RC-PL is the product-limit estimator for left-truncated and right-censored data. The kernel method is the estimator proposed by Eaton et al. (2022). SE is the empirical standard error from $B=500$ bootstraps.
    \end{tablenotes}
    \end{threeparttable}
}
\label{tab:rda_table_wa}
\end{table}

\subsection{Testing results for differences in DemF-RMST with ARIC data}

We presented the results for testing the differences in DemF-RMST between the ARIC participant subgroups (sex, race, and dichotomized age). Web Table \ref{tab:rda_ztest} presented more comprehensive results, including the estimated DemF-RMST in each subgroup, the differences, and the z-test p-values. Confidence intervals are 95\% bootstrap percentile confidence intervals based of 500 bootstrap samples.

\begin{table}[]
\centering
\caption{Estimated restricted mean dementia-free survival time (DemF-RMST) up to 5 and 20 years post-MI in different patient subgroups, and 95\% Bootstrap percentile confidence intervals were shown with units in years. Z-tests for differences in DemF-RMST using the proposed estimator between subgroups were conducted, and the point estimate of DemF-RMST differences of the first subgroup compared to the latter (i.e., male compared to female, black participants compared to white participants, and participants with first MI $<74$ years old compared to participants with first MI $\geq74$ years old), and z-test p-values were shown.}
\resizebox{\linewidth}{!}{
\begin{tabular}{l|ccc|r}
  \hline
  \hline
    & Male & Female & Difference & p-value \\
    5-year post-MI & 2.78 (2.51, 3.05) & 2.96 (2.67, 3.24) & -0.18 (-0.55, 0.24) & 0.36 \\
    20-year post-MI & 5.34 (4.72, 6.05) & 6.14 (5.46, 6.80) & -0.80 (-1.72, 0.18) & 0.092 \\
    \hline
    & Black & White & Difference & p-value \\
    5-year post-MI & 2.64 (2.12, 3.09) & 2.93 (2.72, 3.12) & -0.29 (-0.86, 0.21) & 0.28 \\
    20-year post-MI & 5.04 (4.08, 6.05) & 5.95 (5.44, 6.44) & -0.91 (-2.02, 0.25) & 0.11 \\
    \hline
    & First MI $<74$ years old & First MI $\geq74$ years old & Difference & p-value \\
    5-year post-MI & 3.32 (2.82, 3.84) & 2.82 (2.61, 3.02) & 0.50 (-0.01, 1.09) & 0.078 \\
    20-year post-MI & 8.20 (6.89, 9.55) & 4.69 (4.20, 5.14) & 3.51 (2.12, 5.02) & $<0.0001$ \\
   \hline
   \hline
\end{tabular}
}
\label{tab:rda_ztest}
\end{table}

\label{lastpage}

\end{document}